\documentclass[11pt,a4paper]{article}

\usepackage{url}
\usepackage{hyperref}
\usepackage{verbatim}

\newcommand{\Intern}[1]{}
\newenvironment{example}{\footnotesize\verbatim}{\endverbatim}

\title{Applications of Quantified Constraint Solving over the Reals\\Bibliography}

\author{Stefan Ratschan}

\begin{document}

\maketitle

%\section{Accomplished}

%\section{Introduction}

Quantified constraints over the reals appear in numerous contexts. Usually existential
quantification occurs when some parameter can be chosen by the user of a system, and
univeral quantification when the exact value of a parameter is either unknown, or when it
occurs in infinitely many, similar versions. 

The following is a list of application areas and publications that contain applications
for solving quantified constraints over the reals. The list is certainly not complete, but
grows as the author encounters new items. Contributions are very welcome!

%\section{Bibliography}

\begin{description}
\item[Electrical Engineering/Electronics: ] \cite{Sturm:99,Senesky:03}
\item[Numerical Analysis: ] \cite{Liska:93,Liska:93a,Liska:95,Liska:96,Hong:97,Hong:97c,Liska:07,Vejchodsky:09}
\item[Control: ]
  \cite{Abdallah:96,Jirstrand:96,Jirstrand:97,Dorato:95,Jaulin:96,Nesic:98,Anderson:75,Anai:00b,Glad:95,Anai:99,Champetier:88,Ratschan:01b,Smagina:00,Ratschan:03a,Prakash:02,Nesic:99,Dorato:03,Tibken:00,Dorato:99a,Nguyen:03,Ying:02,Anai:05a,Prajna:05c,Hyodo:07,Wan:08,Vosswinkel:18,Roebenack:18,Roebenack:18a,Vosswinkel:19,Robenack:20},
  survey \cite{Dorato:00},
  reachability~\cite{Lafferriere:99,Lafferriere:99a,Lafferriere:01,Anai:00c}, embedded
  control systems~\cite{Tiwari:02}, hybrid
  systems~\cite{Fraenzle:99,Gao:03,Ghosh:03,Tiwari:02a,Senesky:03,Fotiou:06,Julius:06}, projection of
  system output function~\cite{Pau:06}, computation of control invariant sets~\cite{Bravo:05},
  optimal control~\cite{Pang:2006}, controllability for systems with complex discrete part~\cite{Cimatti:15}, model predictive control~\cite{siaulys:16}, sliding-mode control~\cite{Monnet:18}
\item[Computational Geometry/Motion Planning/Collision Detection: ] \cite{Sturm:98b,Jaulin:96,Hong:91,Weispfenning:01,Arnon:88,Salas:14}
\item[Constraint Databases: ]
  \cite{Kuper:00,Brodsky:96,Paredaens:98,Basu:99,Afrati:94,Baudinet:95,Gaede:95,Koubarakis:94}
\item[Theorem Proving in Real Geometry: ] \cite{Sturm:98a,Dolzmann:99,Brown:00}
  \item[Program Analysis: ] \cite{Colon:03,Cousot:05,groesslinger:04,groesslinger:06,Chen:07,Chen:07a,Xia:09,Erascu:14,Erascu:14a},
\item[Several Different: ] \cite{Dolzmann:97,Weispfenning:97,Loos:93,Jaulin:96}
\item[Use of Predicate Language for Modeling Engineering Problems: ] \cite{Faltings:96,Bowen:93}
\item[Other: ] camera motion: \cite{Benhamou:00}, constraint logic programming:
  \cite{Hong:92}, mechanical engineering: \cite{Ioakimidis:99a,Ioakimidis:97,Ioakimidis:17,Ioakimidis:17a,Ioakimidis:17b,Ioakimidis:17c,Ioakimidis:23} and further papers by Nikolaos Ioakimidis, mathematics:
  \cite{Lazard:88} (from~\cite{Kahan:75}), Hopf bifurcations~\cite{ElKahoui:00,Errami:15}, biology: \cite{Chauvin:94,Ghosh:03,Piazza:05,Brown:06,Anai:06,Yoshida:07,Yoshida:08,Campagna:09,Hong:13,Bradford:19}, epidemiology~\cite{Udovicic:24}, interpolation:~\cite{GonzalezVega:96a}, scheduling~\cite{Gerber:95},
  automated theorem proving~\cite{Bjorner:97,Janicic:99,Janicic:01, Neumaier:07},
  optimization:~\cite{Anai:98}, analysis of particle swarm optimization~\cite{Gerwien:21}, termination of rewrite
  systems~\cite{Collins:91,Dershowitz:87,Lucas:07}, flight control~\cite{Geser:02},  hybrid
  systems~\cite{Prajna:03}, injectivity test~\cite{Braems:01} (see Lagrange/Delanoue/Jaulin
  papers), computer assisted proofs~\cite{Csendes:07,Zwick:02}, parameter estimation~\cite{Braems:00,Braems:05,Kieffer:05}, robotics~\cite{Viegas:17}, economics~\cite{Mulligan:18}, model checking~\cite{Ren:18}, neural network analysis~\cite{Ren:19}, dynamical systems~\cite{Roebenack:21}, normal cone computation~\cite{Mandlmayr:24}, explored area of line-sweep sensor~\cite{Costa:24} (the paper uses a costum algorithm to solve a quantified problem), formal modeling using Event-B~\cite{Kobayashi:24}

\end{description}

\section{Acknowledgments}

%This work has been supported by a Marie Curie fellowship of the European Union under
%contract number HPMF-CT-2001-01255.

Thanks to Hirokazu Anai for contributing references.

\bibliographystyle{abbrv}

\bibliography{sratscha}

%\appendix

%\begin{ifhtml}
\section{Examples}

%checked papers:
%Abdallah:96: first example trivial, second example not not explicitely given (we
% might compute it), third example included, fourth example included (also in
% \cite{Dorato:95})
%Afrati:94: no interesting examples
%Anai:98: do not have it
%Anai:00b: do not have it
%Anai:00c: Available from a techrep, but complicated to type in, ask Anai!
%Anai:99: do not have it
%Anderson:75: do not have it
%Barmish:90: one example included, other?
%Basu:99: No examples
%Benhamou:00a: ask Frederic
%Bjorner:97: no examples
%Bowen:93: no non-trivial examples
%Brodsky:96: no non-trivial examples
%Brown:00: Continue on this (see example/appFOC/Ellipse-1)
%Champetier:88: too easy
%Chauvin:94: ongoing, not all examples extracted

%promising examples:
%Geser:02: see ~/research/examples/tiwari/, and Tiwari's mail

A collection of examples from the above papers. I am thankful for any contribution!

\begin{itemize}
\item Robust-1~\cite{Dorato:00}:

\begin{example}
FORALL([p], [[0, 1]], 
  9 + 48 p + 48 q + 32 p q > 0 /\
  1 + p + q > 0 /\
  -16 p - 16 q + 16 p^2 + 16 q^2 + 7 > 0 
);
[q];
[[-2, 2]];
\end{example}

Solution:~\cite{Dorato:00}: $4q-1<0 /\backslash 16q+3>0 /\backslash 4q-3>0$, that is $(-0.1875, 0.25)$ union $(0.75, \infty)$.

\item Robust-2~\cite{Fiorio:93,Malan:97}:

\begin{example}
FORALL([q1, q2, ww], [[0.8, 1.25], [0.8, 1.25], [0.0, infty]],
 -k2 - k1 q1 > 0 /\
 - q1 k1 - 50 > 0 /\
 2 k2^2 ww^2  + 4 k2^2 ww k1 q1 q2 + k2^2 k1^2 q1^2 q2^2 + 2 k2^2 ww q2^2 + 4 k2 ww k1 q1 q2^2 + ww k1^2 q1^2 q2^2 > 0 /\
 400 k2^2 ww^2 + 800 k2^2 ww k1 q1 q2 + 400 k2^2 k1^2 q1^2 q2^2 + 400 k2^2 ww q2^2 + 800 k2 ww k1 q1 q2^2 + 400 ww k1^2 q1^2 q2^2 - k2^2 k1^2 q2^2 - ww^2 k1^2 - k2^2 ww k1^2 - ww k1^2 q2^2 > 0
);
[k1, k2];
[[-200, 0], [0, 10]];
\end{example}

\item Robust-3~\cite{Dorato:95}:

\begin{example}
FORALL([ p1, p2 ], [ [0.8, 1.25], [0.8, 1.25]],
  p2*(1+p1*q1) < 0 /\
  FORALL([w1], [[0,10]], 99*w1^2 + p2^2 * (100*(1+p1*q1)^2 - 1) > 0) /\
  FORALL([w2], [[-infty,infty]], (400-q1^2)*w2^2 + p2^2*(400*(1+p1*q1)^2-q1^2) > 0)
);
[q1];
[ [-50, 10] ];
\end{example}

Solution~\cite{Dorato:95}: $-20 \leq q_1 < -1.375$

\item Robust-4~\cite{Malan:97,Barmish:90}

\begin{example}
FORALL([p1, p2, p3, p4], [[2.5, 3.5], [1.5, 2.5], [2.5, 3.5], [9.5, 10.5]],
  p2 q2 + p4 q4 > 0 /\
  p4 + q4 + p1 q1 - p3 q3 > 0 /\
  p1 p3 q2 q4 - p1 p3 q2 p4 - p2 q1 p3 p4 - 2 p1 p2 q1 q2 - p1 q2 p4 q3 + p2 q1 p3 q4 - p2 q1 p4 q3 + 2 p2 p3 q2 q3 + p1 q2 q3 q4 + p2 q1 q3 q4 + 2 p3 p4 q3 q4 - p1 q1 p3 p4 q3 - p1 q1 p3 q3 q4 - p2 p3^2 q2 - p2 q2 q3^2 - p3 p4^2 q3 - p3 p4 q3^3 - p3 q3 q4^2 - p3^3 q3 q4 - p1 p2 q1^2 p3 + p1 p2 q1^2 q3 - p1^2 q1 p3 q2 + p1^2 q1 q2 q3 + p1 q1 p4 q3^2 + p1 q1 p3^2 q4 - p1 p3 q2 q3^2 + p1 p3^2 q2 q3 - p2 q1 p3 q3^2 + p2 q1 p3^2 q3 - p1^2 q2^2 - p2^2 q1^2 + p3^2 p4 q3^2 + p3^2 q3^2 q4 > 0 /\
  q3 - p3 > 0    
);
[q1, q2, q3, q4];
[[19, 21], [22, 24], [9, 11], [4, 5]];
\end{example}

\item Robust-5~\cite{Jaulin:96}

\begin{example}
FORALL([z, T, w0, K], [[0.95, 1.05], [-1.05, -0.95], [0.95, 1.05], [0.95, 1.05]],
  K c1 w0^2 < 0 /\
  2 z T w0 + 1 < 0 /\
  (2 z w0 + w0^2 T + w0^2 T c3)*(2 z T w0 + 1) - w0^2 (K*c2+1) < 0 /\
  w0^2*(K*c2 + 1)*(2*T*z*w0 + 1)*(2*z*w0 + w0^2*(T + K*c3)) - K*c1*w0^2*(2*T*z*w0 + 1)^2 - w0^4*(K*c2 + 1)^2 < 0
);
[ c1, c2, c3 ]; 
[[ -10, 10], [-10, 10], [-10, 10]];
\end{example}

\item Robust-6~\cite{Jaulin:02}

\begin{example}
FORALL([p1, p2, p3], [[0.9, 1.1], [0.9, 1.1], [0.9, 1.1]],
  (1 + c2*p1 )*( (p2*p3^2 +p3)*(p2*p3 +1) - p2*(p3^2 +c2*p1*p3^2)) - (p2*p3 + 1)^2*(c1*p1) > 0 /\
  (p2*p3 + 1)*(p2*p3 + 1) - p2*( p3 + c2*p1*p3) > 0 /\
  c1*p1*p3^2 > 0
);
[c1, c2];
[[0, 1], [0, 1]];
\end{example}

\item Control-Stabilization-1~\cite{Abdallah:96}

\begin{example}
EXISTS([A, B, D, P1, P2, P3, P4, P5, P6, P7, P8, P9],
  P1 > 0 /\ P2 > 0 /\ P3 > 0 /\ P4 P5 > 0 /\ A > 0 /\ B P6 P7 > 0 /\ P8 P9 > 0 /\
  P1 = A B^2 - D^2 /\
  P2 = - A B + A + D^2 - D - 1 /\
  P3 = A B - A D - 2 A + D^3 + 4 D^2 + 4 D /\
  P4 = A B - 2 A - B D^2 - 4 B D - 4 B + 2 D^2 + 5 D + 2 /\
  P5 = A B^3 - A B^2 D - 4 A B^2 + 2 A B D + 4 A B + 2 B D^3 + 5 B D^2 + 2 B D - D^3 - 4 D^2 - 4 D /\
  P6 = A B - 2 A - B D^2 - 4 B D - 4 B + 2 D^2 + 4 D /\
  P7 = A B^2 - A B D - 4 A B + 2 A D + 4 A + 2 D^3 + 4 D^2 /\
  P8 = A B - 2 A - B D^2 - 4 B D - 4 B + 2 D^2 + 3 D - 2 /\
  P9 = A B^3 - A B^2 D - 4 A B^2 + 2 A B D + 4 A B + 2 B D^3 + 3 B D^2 - 2 B D + D^3 + 4 D^2 + 4 D 
);
\end{example}

Solution: True, witness: A=110, B=3/2, D=15

\item Biology-1~\cite{Chauvin:94}

\begin{example}
FORALL([c, t, r, h, d, v], 
  (r>0 /\ t>0 /\ c<1 /\ d>0 /\ v>0) ==> 
     ((d+1)(v+t+1) = r h (c t + d + 1) ==> -(d+1)(v d + v + t d + t + d + 1 - h r c t - h r d - h r)=0))
\end{example}

Solution~\cite{Chauvin:94}: True  

\item Biology-2~\cite{Chauvin:94}

\begin{example}
FORALL([r2, p1], 
  (r2>1/2 ==> 
    (4 (180 + 180 r2^2) = 75 (12+6 r2) /\
      (8 + 24 p1 + 20 r2 - 16 r2^2 p1 + 198 r2^2 p1^2 + 95 p1 r2 + 185 r2 p1^2 - 32 r2^2 - 32 p1^2) p1 = 0) ==> p1 <= 0 ))
\end{example}
  
Solution~\cite{Chauvin:94}: True
% solvable without singular information!!!

\item Biology-3~\cite{Chauvin:94}

\begin{example}
FORALL([r1, r2, p1],
  (r2>0 /\ r1>0 /\ r2>r1) ==> 
    (( 4 (720 r1^2 + 180 r2^2)= 75 (24 r1 + 6 r2) /\
         (-88 r1 r2^2 p1^2 + 56 r1 r2^2 p1 - 480 r1^2 p1^2 r2 - 335 r1 p1 r2 + 55 r2 r1
  p1^2 + 480 r1^2 p1 r2 - 80 r1^2 + 128 r1^3 + 80 r1^2 p1 - 20 r2^2 p1 - 20 r2 p1 - 55
  r2^2 p1^2 - 256 r1^3 p1 + 128 r1^3 p1^2 + 32 r1 r2^2) p1 = 0) ==> p1 <= 0)
\end{example}

Solution~\cite{Chauvin:94}: True

\item Biology-4~\cite{Chauvin:94}

\begin{example}
FORALL([g2, p1], 
  (r2>0 /\ r1>0 /\ r2>r1 /\ 0<g2 /\ g2 < 1/2) ==> 
    (( 4 (900 r1^2 (1-g2) + 900 r2^2 g2 ) = 75 (30 r1 (1-g2) + 30 r2 g2) /\
        (55 r2^2 g2 p1^2 + 20 r2^2 g2 p1 + 20 r2 g2 r1 - 40 r2 g2 r1 p1 + 20 r1^2 - 20
  r1^2 g2 - 32 r1^3 + 32 r1^3 g2 + 75 r1 p1 r2 - 55 r2 g2 r1 p1^2 - 120 r1^2 p1 r2 + 120
  r1^2 p1 r2 g2 + 120 r1^2 p1^2 r2 - 120 r1^2 p1^2 r2 g2 - 56 r1 r2^2 g2 p1 + 88 r1 r2^2
  g2 p1^2 - 32 r1^3 p1^2 + 32 r1^3 p1^2 g2 - 20 r1^2 p1 + 20 r1^2 p1 g2 + 64 r1^3 p1 -
  64 r1^3 p1 g2 - 32 r1 r2^2 g2) p1 = 0 ) ==> p1 <= 0))
\end{example}

Solution~\cite{Chauvin:94}: True

\item Biology-5~\cite{Chauvin:94}

\begin{example}
FORALL([p1], (t > 0) ==> (( 2 (48 t + 7) = 7 (1 + 8 t) /\
  -3472875 p1 - 26162325 p1^2 + 10584000 t - 69325200 t p1 - 327499200 t^2 p1 - 578283300
t p1^2 - 4222108800 t^2 p1^2 - 10163232000 t^3 p1^2 + 1852200 = 0) ==> (p1 <= 0)))
\end{example}

Solution~\cite{Chauvin:94}: True

\item Biology-6~\cite{Chauvin:94}

\begin{example}
FORALL([p1], (t > 0 /\ 0 < c /\ c < 1) ==> (( 2 (60 c t + 7) = 7 (1 + 8 t) /\
  139179600 t p1 - 87318000 c t p1 - 662256000 c t^2 p1 + 357210000 t p1^2 c + 5765256000
  t^2 p1^2 c + 3472875 p1 + 14817600 t + 26162325 p1^2 + 23250240000 c t^3 p1^2 +
  292515300 t p1^2 + 857304000 t^2 p1 - 390096000 t^2 p1^2 - 31752000 c t - 8436960000 t^3
  p1^2 - 1852200 = 0) ==> (p1 <= 0))
\end{example}

Solution~\cite{Chauvin:94}: True

\item Biology-7~\cite{Chauvin:94}

\begin{example}
FORALL([p1], (t > 0 /\ 2 r2 > 1) ==> (( 8 (1+r2^2) (1+ 6 t) = (2 + r2) 5 (1 + 8 t) /\
  -370440 - 1111320 p1 + 8678880 r2^2 t p1 - 926100 r2 + 25401600 r2^2 t^2 p1 - 3538080000
r2^2 t^3 p1^2 + 1481760 p1^2 + 29529360 t p1^2 - 91551600 t p1 r2 - 153071100 t p1^2 r2 -
791985600 t^2 p1^2 r2 - 202127940 r2^2 t p1^2 - 1472385600 r2^2 t^2 p1^2 - 5927040 t +
740880 r2^2 p1 - 4398975 p1 r2 - 7726320 t p1 + 182347200 t^2 p1^2 - 9168390 r2^2 p1^2 -
8566425 r2 p1^2 + 8890560 r2^2 t + 1481760 r2^2 - 450878400 t^2 p1 r2 - 925344000 t^3 p1^2
r2 - 7408800 t r2 + 12700800 t^2 p1 + 326592000 t^3 p1^2 = 0) ==> (p1 <= 0)))
\end{example}

Solution~\cite{Chauvin:94}: True

\item Control-1~\cite{Jirstrand:97}, Example 2.2

\begin{example}
EXISTS([u], [[-0.5, 0.5]],
  -x1 + x2 u = 0 /\ -x2 + (1+x1^2) u + u^3 = 0 
)
\end{example}

Solution~\cite{Jirstrand:97}: 
\begin{example}
x2^4 - x1^3 x2^2 - x1 x2^2 - x1^3 = 0 /\ (x2 + 2 x1 <= 0 \/ x2 - 2 x1 >= 0)
\end{example}

\item Control-2~\cite{Jirstrand:97}, Example 4.1

\begin{example}
FORALL([t], [[0, 1]]
  EXISTS([u, l], [[-1,1], (0, \infty]],
    -t + 2 = l /\ -(3 t^2 - 2 t^3) - t^2 + 4 u = l ( 6 t - 6 t^2 ) ))
\end{example}

\item Reachability-1~\cite{Lafferriere:01}, Example 3.4

\begin{example}
EXISTS([z], [[ 1, \infty ]],
  x2 >= 3 /\
  x1 z^2 = 4 /\
  (2 x2 - 1) + z^2 = 6 z
)
\end{example}

Solution: $x_2\geq 3 \wedge 4 x1^2 x2^2 - 4 x1^2 x2 + 16 x1 x2 + x1^2 - 152 x1 + 16 = 0$

\item Reachability-2~\cite{Lafferriere:01}, Example 3.5

\begin{example}
EXISTS([a, z], [[ 0, 1 ], [ 1, \infty ]],
  y1 = 2/3 a (-z^4 + z) /\
  y2 z^2 = 1/2 a (z^4 - 1)
)
\end{example}

\begin{example}
EXISTS([z], [[ 1, \infty ]],
  3 y1 (z^3 + z^2 + z + 1) + 4 y2 (z^5 + z^4 + z^3) = 0
)
\end{example}

Solution: $(y_2>0 \wedge y_1+y_2\leq 0) \vee 
           (y_2<0\wedge y_1+y_2\geq 0) \vee
           4 y_2 + 3 y_1 = 0$

\item Reachability-3~\cite{Lafferriere:01}, Example 3.6

\begin{example}
EXISTS([w, z], [[-2,2], [-2,2]],
  w^2+z^2 = 1 /\
  0 = (z^2-w^2)-2/3 w /\
  y2 = - 4 z w - 5/3 z
);
[y2];
[[0, 4]];
\end{example}

Solution: $y2=\frac{\sqrt{-181\sqrt{19}+908}}{9\sqrt{2}} \vee
y2=\frac{\sqrt{181\sqrt{19}+908}}{9\sqrt{2}}$

\item Reachability-4~\cite{Lafferriere:01}, Example 3.7

\begin{example}
EXISTS([w, z, a], [[-2,2], [-2,2], [0, \infty]],
   a>0 /\
   w^2+z^2 = 1 /\
   -3a = w ((4 a^2 - 2) z + 2 - a^2) /\
   3a = (a^2-2)(w^2-z^2+z)
);
[];
[];

\end{example}

\end{itemize}

%\end{ifhtml}

\end{document}